# Nonlinear evolution of weak discontinuity waves in Darcy-type porous media

Mithilesh Singh[*]
Department of Applied Science, Rajkiya Engineering College, Sonbhadra-231206, INDIA

## Abstract

The propagation of nonlinear waves in one dimensional space, unsteady and compressible flow in Darcy-type porous media is analyzed. It is assumed that the weak discontinuity propagate long the characteristic path using the characteristics of the governing quasilinear system as the reference coordinate system. Evolution equation in the characteristic plane is derived. As an application of the theory the breaking point at the wave front is determined. It is assessed as to how the porosity of the medium affects the process of steepening and flattering of acceleration waves with planar, cylindrical, and spherical symmetry. The critical amplitude of the initial disturbance has been determined such that any compressive disturbance with initial amplitude greater than the critical one always grows into a shock wave, while the initial amplitude less than the critical one always decays.

## 1. Introduction

The propagation of nonlinear travelling waves in porous media has received considerable attention, during the past few decades, due to its applications in geophysics, mining, oil exploration, etc. The problems of linear wave propagation in porous media have been analyzed in "Morse and Ingard (1968)" and "Pascal (1986)". In gas dynamics, the system of Euler's equations representing the conservation of mass, momentum and energy for perfect gas in Darcy-type porous media are used to analyze the problem of nonlinear waves propagation "Nield and Bejan (1999)". Growth and decay of acceleration waves in porous media have been studied by Jordon "Jordan (2005)". The properties of compressible gas flow in a porous have been examined in "Ville (1996)". "Hsiao (1999)" analyzed the initial value problem for the system of compressible adiabatic flow through porous media in the one space dimension with fixed boundary condition. "Pan (2006)" conjectured that Darcy's law governs the motion of compressible porous media flow in large time adiabatic flow. "Jeffrey (1976)" studied the quasilinear hyperbolic systems and wave. "Boillat and Ruggeri (1965)" developed the interaction between a shock and acceleration waves in the ideal gas. The evolution of weak discontinuity waves in self-similar flow and secondary shock formation with point explosion model have been analyzed by Virgopia and Ferraioli in "Virgopia and Ferraioli (1982)". "Ram (1978)" studied the growth and decay of acceleration waves in radiating gas. "Pandey and Sharma (2007)" analyzed the characteristic shock with weak discontinuity in non-ideal gas. A. "Mentrelli et al. (2008)" examined the shock and an accelerations wave in a perfect gas for increasing shock strength. "Singh et al. (2010)" have been analyzed the propagation of nonlinear travelling waves in Darcy-type porous by wave front expansion technique.

---
[*] Email- msingh.rs.apm@itbhu.ac.in

In the present article, the characteristic coordinates are used to analyze the nonlinear wave's propagation in a fluid that saturates a fixed, rigid, homogeneous and isotropic porous media. The effect of pores in the flow region has been incorporated using the Darcy law given as follows "Nield and Bejan (1999)"

$$\nabla P = -(\mu\chi/K)u, \qquad (1.1)$$

where $P$ is the intrinsic pressure, $u$ is the velocity vector, $K$, $\mu$, and $\chi(<1)$ are the positive constants denote the permeability, dynamic viscosity and porosity, respectively. Also, the filtration velocity vector $\upsilon$ is related to the intrinsic velocity vector $u$ by the Dupuit-Forchheimer relationship $\upsilon = \chi u$. In fact, Darcy law is equivalent to a body force term.

To the best of our knowledge, such an analytical description of a complete history of acceleration waves has not been studied previously in Darcy-type porous media. "Ram (1978)", "Boillat and Ruggeri (1965)" have developed a theory for one-dimensional satisfying the Euler equations, in order to achieve a detailed comprehension of the consequences of the compression waves steepen into shock and expansion waves relax. This paper brings out some interesting features of the evolutionary behavior of acceleration waves in a Darcy-type porous media.

## 2. Governing equations

The governing equations for a viscous, compressible and homogeneous fluid flowing in a fixed, rigid, non-thermally conducting and isotropic porous medium of porosity $\chi$ and permeability $K$ can be written as

$$\rho_t + u\rho_r + \rho u_r + \frac{m\rho u}{r} = 0, \tag{2.1}$$

$$\rho(u_t + uu_r) + p_r + \frac{\mu\chi}{K}u = 0, \tag{2.2}$$

$$e_t + ue_x - p\rho^{-2}(\rho_t + u\rho_x) = -\frac{\mu\chi}{K\rho}u^2, \tag{2.3a}$$

where $\rho$ denotes the gas density, $u$ is the velocity, $p$ represents the pressure, $e$ is the internal energy, $\gamma$ denotes the specific heat ratio, $t$ is the time and $x$ is the spatial coordinate. Here $m$ takes values 0, 1 or 2 for planar, cylindrical and spherical symmetry, respectively. The subscripts denote partial differentiation unless stated otherwise.

The internal energy of an ideal gas is given by

$$e = p/\rho(\gamma-1),$$

Using the above relation, Eq. (2.3a) may be written as

$$p_t + up_r - \frac{\gamma p}{\rho}(\rho_t + u\rho_r) + (\gamma-1)\frac{\mu\chi}{K}u^2 = 0. \tag{2.3b}$$

The above system of Eqs. (2.1)-(2.3b) are supplemented with an equation of state $p = \rho RT$, where $R$ is gas constant and $T$ represents the temperature.

The governing system of Eqs. (2.1), (2.2) and (2.3b) can be written in the following matrix form

$$V_t + AV_r + B = 0, \tag{2.4}$$

where

$$V = \begin{bmatrix} \rho \\ u \\ p \end{bmatrix}, A = \begin{bmatrix} u & \rho & 0 \\ 0 & u & 1/\rho \\ 0 & \gamma p & u \end{bmatrix} \text{ and } B = \begin{bmatrix} m\rho u/r \\ \mu\chi u/\rho K \\ \gamma mpu/r + (\gamma-1)\mu\chi u^2/K \end{bmatrix} \tag{2.5}$$

An acceleration wave is a propagating singular surface in order to across the motion, and the velocity and deformation gradient are continuous, but second and higher order derivatives of motion suffer finite jump discontinuity. The jump $[\ddot{r}]$ is taken to be an amplitude vector of the acceleration wave. The function $V(r,t)$ is continuous everywhere except at characteristic curve $\Sigma(t)$, but $V_r$ and $V_t$ have finite jumps and it is said to be a weak discontinuity at this curve.

Furthermore, the matrix $A$ has three families of characteristic curves $C_0, C_-$ and $C_+$ respectively

$$\frac{dr}{dt} = u, \tag{2.6}$$

$$\frac{dr}{dt} = u - a \quad \text{and} \quad \frac{dr}{dt} = u + a, \tag{2.7a, b}$$

where $a = (\gamma p/\rho)^{1/2}$ is the sound speed. For isentropic flow, the speed of sound is given by "Courant and Friedrichs (1948)".

$$a^2 = (\partial p/\partial \rho)_{s=const.}. \tag{2.8}$$

where $s$ is the entropy which is constant along the particle's path.

## 3. Evaluation of the weak discontinuity

To study the nonlinear effects on the wave pattern, we introduce a new coordinate system $(\alpha, \beta)$, where $\alpha$ is constant along the outgoing wave $dr/dt = u + a$ in $(r,t)$ the plane, if it originate at time $\alpha = t'$. $\beta$ is a particle tag so that $\beta$ is constant along the particle path $dr/dt = u$ in $(r,t)$ the plane. If the wave front transverse a particle at an instant $t^*$, these particles and its path are labeled by $\beta = t^*$.

Each pair of values $(\alpha, \beta)$ there is a corresponding pair $(r,t)$ so that $r = r(\alpha, \beta)$, $t = t(\alpha, \beta)$

The view of the nature of $\alpha$ and $\beta$ will be satisfied the following PDE'S

$$r_\alpha = u t_\alpha \quad \text{and} \quad r_\beta = (u + a) t_\beta, \tag{3.1a, b}$$

In consequence of this transformation, we have

$$V_t = \frac{V_\beta r_\alpha - u_\alpha r_\beta}{J} \quad \text{and} \quad V_r = \frac{V_\alpha t_\beta - V_\beta t_\alpha}{J}, \tag{3.2a, b}$$

To ensure that a plane between $(r,t)$ and $(\alpha, \beta)$ are one-to-one correspondence to each other and it is essential that the Jacobian transformation, $J = \partial(r,t)/\partial(\alpha, \beta) = -at_\alpha t_\beta$ is neither zero nor undefined. Since the doubling up or overlapping of fluid particles is prohibited from physical considerations, a breakdown of the solution will arise $J = 0$ if and only if $t_\alpha = 0$, when two adjoining characteristics convert into a shock wave. Hence, $J = 0$ will also give us the condition for the steepening of the wave front into the shock wave which is given "Courant and Friedrichs (1948)".

Using Eq. (3.2) in (2.1), (2.2) and (2.3b) yield

$$at_\alpha \rho_\alpha - \rho t_\beta u_\alpha + \rho u_\beta t_\alpha + \frac{m\rho u t_\alpha t_\beta}{r(\alpha, \beta)} = 0, \tag{3.3}$$

$$\rho a u_\alpha t_\beta - t_\beta p_\alpha + p_\beta t_\alpha + \frac{\mu \chi u}{K} = 0, \tag{3.4}$$

$$ap_\alpha t_\beta - \gamma p \left( t_\beta u_\alpha - t_\alpha u_\beta + \frac{m a u t_\alpha t_\beta}{r(\alpha, \beta)} \right) + \frac{\gamma m p u}{r} + \frac{(\gamma - 1)\mu \chi}{K} u^2 = 0. \tag{3.5}$$

Using Eqs. (3.4) and (3.5), (3.3) may be written as

$$p_\beta + \rho a u_\beta + \frac{m \rho a^2 u t_\beta}{r(\alpha, \beta)} + \frac{\mu \chi u}{K}(1 + (\gamma - 1)u) + \frac{\gamma m p u}{r(\alpha, \beta)} = 0. \tag{3.6}$$

Across a characteristic shock, the Rankine-Hugoniot conditions are given as

$$[u] = 0, \quad [\rho] = 0, \quad [p] = 0, \quad t = \beta \quad \text{at} \quad \alpha = 0. \tag{3.7}$$

Further, the gas is homogeneous and at rest ahead of the wave front, Eq. (3.7) is written as

$$u_\beta = 0, \quad p_\beta = 0, \quad \rho_\beta = 0, \quad t_\beta = 1 \quad \text{at} \quad \alpha = 0, \tag{3.8}$$

On using the Eqs. (3.7)- (3.8) in (3.4) at the wave front, we get

$$p_\alpha = a_0 \rho_0 u_\alpha \quad \text{at} \quad \alpha = 0. \tag{3.9}$$

Eq. (3.1a, b) at the wave front may be written as

$$r_\beta = a_0, \quad r_\alpha = 0 \quad at \quad \alpha = 0. \tag{3.10}$$

To determine the amplitude of the acceleration wave $\Pi = [u_r]$ at the wave front from Eqs. (3.1b) and (3.8), we get

$$[u_r] = \Pi = -u_\alpha / at_\alpha \neq 0 \quad \text{at} \quad \alpha = 0, \tag{3.11}$$

Differentiating (3.6), (3.9) and (3.1a, b) with respect to $\alpha$ and $\beta$ we have

$$\frac{u_{\alpha\beta}}{t_\alpha} = \frac{a_0}{2\rho_0}\left[\frac{\mu\chi}{K} + \frac{m\rho_0}{\beta}\right]\pi \quad \text{at} \quad \alpha = 0, \tag{3.12}$$

$$\frac{t_{\alpha\beta}}{t_\alpha} = \frac{\gamma+1}{2}\pi \quad \text{at} \quad \alpha = 0, \tag{3.13}$$

Differentiating of Eq. (3.11) with respect to $\beta$, and using it in (3.12)-(3.13) we get

$$\frac{d\Pi}{d\beta} + \frac{1}{2}\left(\frac{\nu\chi}{K} + \frac{m}{\beta}\right)\Pi + \frac{\gamma+1}{2}\Pi^2 = 0 \quad \text{at} \quad \alpha = 0, \tag{3.14}$$

where $\nu = \mu/\rho_0$ is the kinematic viscosity.

Introducing the non-dimensional variables $\zeta$, $\tau$ and $\Omega$ in following form

$$\zeta = \Pi/\Pi^*, \quad \tau = (\beta - \beta^*)/2\beta^*, \quad \Omega = (\gamma+1)\Pi^*\beta^*, \tag{3.15}$$

where $\Pi^*$ and $\beta^*$ are parameters.

Substituting (3.15) in (3.14) we get the first order nonlinear differential equation (Riccati type equation) in the following form

$$\frac{d\zeta}{d\tau} + \left(\frac{\nu\chi}{2K} + \frac{m}{1+2\tau}\right)\zeta + \Omega\zeta^2 = 0, \tag{3.16}$$

Eq. (3.16) is a nonlinear ODE; it can be easily transformed to a linear ODE in $1/\zeta$ by dividing with $\zeta^2$:

$$\frac{d}{d\tau}\left(\frac{1}{\zeta}\right) - \left(\frac{\nu\chi}{2K} + \frac{m}{1+2\tau}\right)\frac{1}{\zeta} = \Omega, \tag{3.17}$$

Eq. (3.17) is a first-order linear ODE whose solution is given by

$$\zeta = \frac{\phi(\tau)}{(1+\Omega\, Q(\tau))}, \tag{3.18}$$

where $Q(\tau) = \int_0^\tau \frac{e^{-\nu\chi\tau/2K}}{(1+2\tau)^{m/2}}d\tau$ and $\phi(\tau) = \frac{e^{-\nu\chi\tau/2K}}{(1+2\tau)^{m/2}}$.

In Eq. (3.18), the coefficient of Darcy term has become $\delta = \chi/\text{Re}$, where is $\text{Re} = a_0(0)l\rho_0/\mu$, we defined $l = K/L$ for convenience. We also note that the continuum assumption demands $K_n < 0.01$, where $K_n = \sqrt{K}/L$ is Knudsen number, $\sqrt{K}$ is the average molecular free path, $L$ is physical length scale and $l$ is characteristic length scales "(Nield and Bejan (1999))" and "Singh et al. (2010)". $\Omega$ is known as the initial amplitude of acceleration wave, $\zeta \neq 0$ denotes the value of $\zeta$ at $\alpha = 0$. The function $\phi(\tau)$ is non-zero, finite and continuous on $[0,\tau)$ and it tends to zero as $\tau \to \infty$ with $Q(\tau) < \infty$. If $\zeta(\tau) \to 0$ as $\tau \to \infty$, is implying the decays of expansion waves. If $\Omega < 0$ is implying the compression waves, it follows from (27) that there are three conditions in "Boillat, and Ruggeri(1965)":

**(1)** Let $|\Omega| < \Omega_c$, where $\Omega_c = 1/\phi(\infty)$ is the critical amplitude of acceleration wave. It is nonzero and continuous in $\tau \in [0,\infty)$, if $\tau \to \infty$ then $\Omega \to 0$ and $\phi(\tau)$ becomes zero, then the decay of the wave front is the decay an expansion wave.

**(2)** Let $|\Omega| = \Omega_c$. Then $\zeta$ is finite, nonzero and continuous in the $[0,\infty]$. Again, $\zeta \to 0$ as $\tau \to \infty$ is implying that wave decays which illustrated through the fig. 1, fig. 2 and fig. 3.

**(3)** Let $|\Omega| > \Omega_c$, then $\phi(\tau_c) = 1/\Omega$ in finite time $\tau_c > 0$. $\zeta$ is finite, nonzero and continuous on $[0,\infty)$ and $|\zeta| \to \infty$ as $\tau \to \tau_c$. This means that compression waves steepen into the shock wave in a finite time $\tau = \tau_c$, only when initial discontinuity increases a critical value.

## 4. Results and discussion

In this paper, we have analyzed with evolutionary behavior of acceleration waves in Darcy-type porous media. Computations of variables $\zeta$ and $\tau$, associated with acceleration waves, have been carried out for plane $(m=0)$, cylindrical symmetrical $(m=1)$ and spherical symmetrical $(m=2)$ flows by taking $\nu = 0.00001496$, $\chi = 0.476$, $K_n = 0.00421$, $L = 1$ and values of $\Omega = -2$, $\Omega = -1$ and $\Omega = 1$,. The steepening of compression wave is slowed down in case of porous medium as compared to the nonporous case in the planer, cylindrical and spherical symmetric flow which is illustrated through the figs. 1- 3. It may also be noted that the compression waves steepen into a shock only if the magnitude of the initial slope of the wave front is greater than a critical value; an expansion wave is relaxed in the entire above planes which is also shown in figs 1-3.

## 5.  Conclusions

In the present article, we have used the characteristics of the governing quasi-linear system as the reference coordinate system to determine an exact solution of the equations that describe an unsteady planar, cylindrical and spherical symmetric flow of Darcy-type porous media. The evolution of weak discontinuity in a state characterized by the exact solution is obtained. The analyses show that the rate of steepening of compression wave into a shock wave of Darcy-type porous media is slowed down as compared to ideal gases and expansion wave is relaxed in both an ideal gas and Darcy-type porous media.

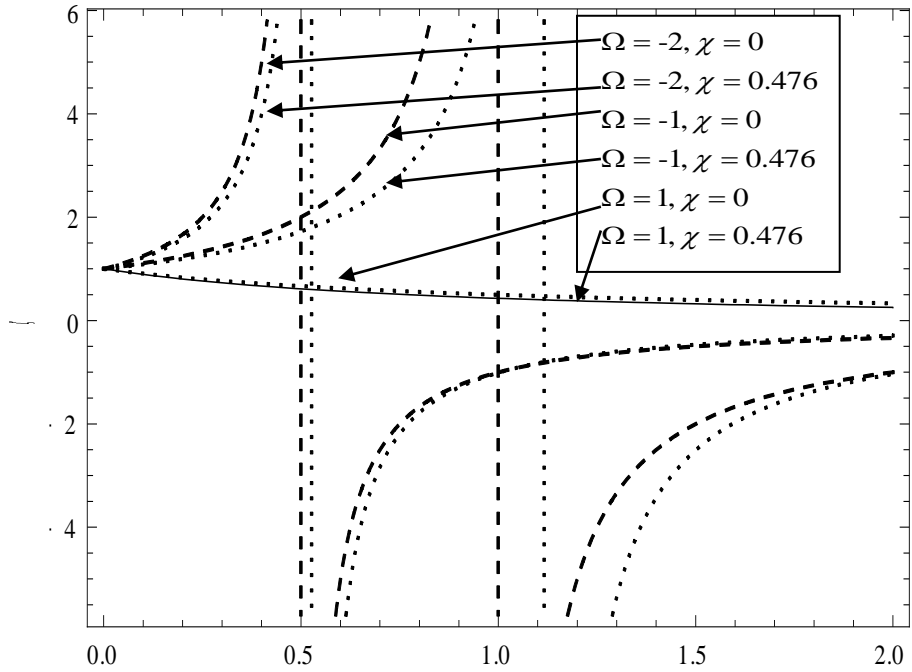

**Fig.1** Evolution of the amplitude of acceleration waves influenced by Darcy-type and non-Darcy type porous media for plane ($m = 0$) flows.

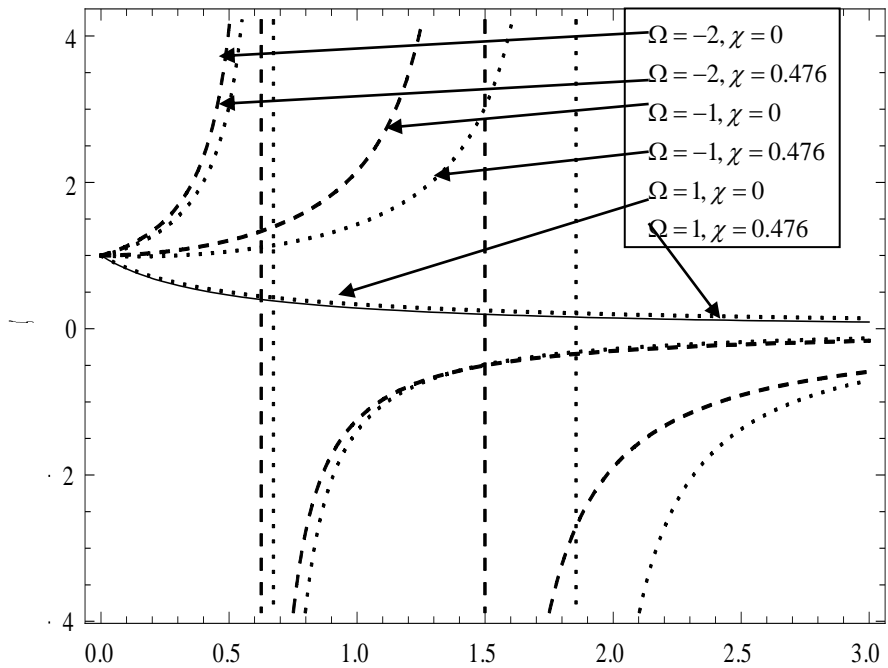

**Fig. 2** Evolution of the amplitude of acceleration waves influenced by Darcy-type and non-Darcy type porous media for cylindrical ($m = 1$) flows.

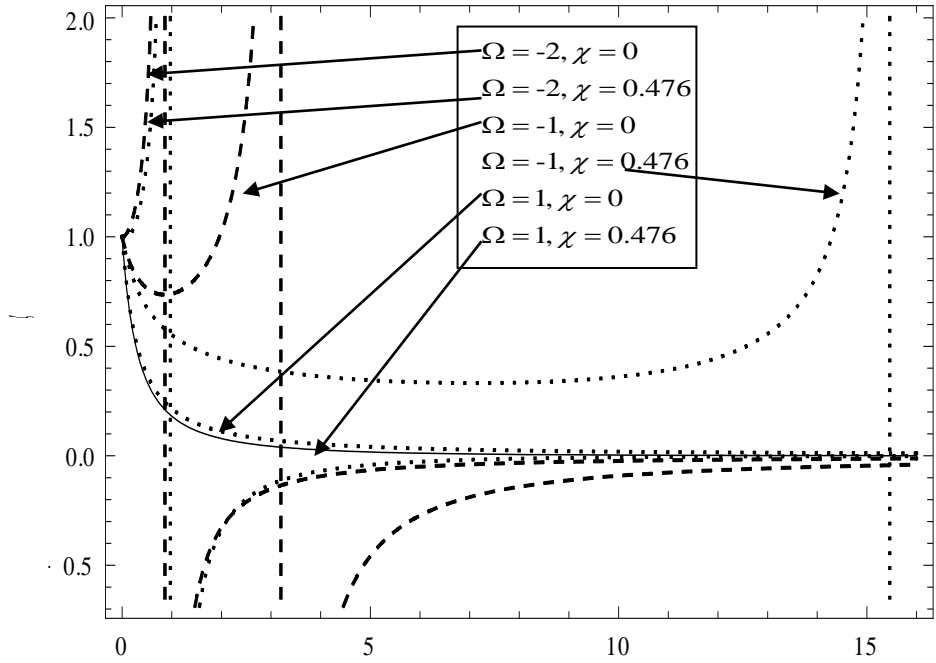

**Fig. 3** Evolution of the amplitude of acceleration waves influenced by Darcy-type and non-Darcy type porous media for spherical ($m=2$) flows.